\documentclass[final,5p,times,twocolumn]{elsarticle}

\usepackage{amssymb}
\usepackage{amsmath}
\usepackage{textcomp}
\journal{Nuclear Instruments and Methods in Physics Research A }
\begin{document}
\begin{frontmatter}
\title{AWAKE, The Advanced Proton Driven Plasma Wakefield Acceleration Experiment at CERN}
\author[cern]{E.~Gschwendtner} %
 \author[oslo]{E.~Adli} %
 \author[ist]{L.~Amorim} %
 \author[cock,lanc]{R.~Apsimon} %
 \author[desy]{R.~Assmann} %
 \author[mpp]{A.-M.~Bachmann} %
 \author[mpp]{F.~Batsch} %
 \author[cern]{J.~Bauche} %
 \author[oslo]{V.K.~Berglyd~Olsen} %
 \author[cern]{M.~Bernardini} %
 \author[ral]{R.~Bingham} %
 \author[cern,czech]{B.~Biskup} %
 \author[cern]{T.~Bohl} %
 \author[cern]{C.~Bracco} %
 \author[jai,ox]{P.~N.~Burrows} %
 \author[cock]{G.~Burt} 
 \author[mpip]{B.~Buttensch\"{o}n} %
 \author[cern]{A.~Butterworth} %
 \author[mpp]{A.~Caldwell} %
\author[ucl]{M.~Cascella} %
 \author[cern]{E.~Chevallay} %
\author[strath]{S.~Cipiccia} %
 \author[cern]{H.~Damerau} %
 \author[ucl]{L.~Deacon}  %
 \author[triumf]{P.~Dirksen} %
 \author[cern]{S.~Doebert} %
 \author[desy]{U.~Dorda} %
 \author[duss]{J.~Farmer} %
\author[cern]{V.~Fedosseev} %
 \author[cern]{E.~Feldbaumer} %
 \author[cock,liv]{R.~Fiorito} %
 \author[ist]{R.~Fonseca} %
 \author[cern]{F.~Friebel} %
  \author[binp,nsu]{A.A.~Gorn} %
 \author[mpip]{O.~Grulke} %
 \author[cern]{J.~Hansen} %
 \author[cern]{C.~Hessler} %
 \author[cern]{W.~Hofle} %
 \author[jai,ox]{J.~Holloway} %
 \author[mpp,tum]{M.~H\"{u}ther}  %
 \author[strath]{D.~Jaroszynski} %
 \author[cern]{L.~Jensen} %
 \author[ucl]{S.~Jolly} %
 \author[mpp]{A.~Joulaei} %
 \author[jai,ox]{M.~Kasim} %
 \author[ucl]{F.~Keeble} %
  \author[cock,manc]{Y.~Li}  %
 \author[triumf]{S.~Liu} %
 \author[ic,ist]{N.~Lopes} %
 \author[binp,nsu]{K.V.~Lotov} %
 \author[ucl]{S.~Mandry} %
  \author[duss]{R.~Martorelli} %
 \author[mpp]{M.~Martyanov} %
 \author[cern]{S.~Mazzoni} %
 \author[cock,manc]{O.~Mete} %
 \author[binp,nsu]{V.A.~Minakov} %
 \author[cock,lanc]{J.~Mitchell} %
 \author[mpp]{J.~Moody} %
 \author[mpp]{P.~Muggli} %
 \author[ic,jai]{Z.~Najmudin} %
 \author[ox,ral]{P.~Norreys} %
  \author[mpp]{E.~\"{O}z} %
 \author[cern]{A.~Pardons} %
 \author[cern]{K.~Pepitone} %
 \author[cern]{A.~Petrenko} %
 \author[cern,epfl]{G.~Plyushchev} %
 \author[duss]{A.~Pukhov} %
 \author[mpp,tum]{K.~Rieger} %
 \author[lmu]{H.~Ruhl} %
 \author[cern]{F.~Salveter} %
 \author[mpp,triumf,uv]{N.~Savard} %
 \author[cern]{J.~Schmidt} %
 \author[jai,ox]{A.~Seryi} %
 \author[cern]{E.~Shaposhnikova} %
 \author[strath]{Z.M.~Sheng} %
 \author[ucl]{P.~Sherwood} %
 \author[ist]{L.~Silva} %
  \author[cern]{L.~Soby} %
 \author[binp,nsu]{A.P.~Sosedkin} %
 \author[binp,nsu]{R.I.~Spitsyn} %
 \author[ral]{R.~Trines} %
 \author[binp,nsu]{P.V.~Tuev} %
 \author[cern]{M.~Turner} %
 \author[triumf]{V.~Verzilov} %
 \author[ist]{J.~Vieira} %
 \author[cern]{H.~Vincke} %
 \author[cock,liv]{Y.~Wei} %
 \author[cock,liv]{C.P.~Welsch} %
 \author[ucl,desy]{M.~Wing} %
 \author[cock,manc]{G.~Xia} %
\author[cock,liv]{H.~Zhang} %

 \address[binp]{Budker Institute of Nuclear Physics SB RAS, 630090, Novosibirsk, Russia}
 \address[cern]{CERN, Geneva, Switzerland}
 \address[cock]{Cockcroft Institute, Warrington WA4 4AD, UK}
 \address[czech]{Czech Technical University, Zikova 1903/4, 166 36 Praha 6, Czech Republic}
 \address[desy]{DESY, Notkestrasse 85, 22607 Hamburg, Germany}
 \address[duss]{Heinrich-Heine-University of D\"{u}sseldorf, Moorenstrasse 5, 40225 D\"{u}sseldorf, Germany}
 \address[ist]{GoLP/Instituto de Plasmas e Fus\~{a}o Nuclear, Instituto Superior T\'{e}cnico, Universidade de Lisboa, Lisbon, Portugal}
 \address[ic]{John Adams Institute for Accelerator Science, Blackett Laboratory, Imperial College London, London SW7 2BW, UK}
 \address[jai]{John Adams Institute for Accelerator Science, Oxford, UK}
 \address[lanc]{Lancaster University, Lancaster LA1 4YR, UK}
 \address[lmu]{Ludwig-Maximilians-Universit\"{a}t, 80539 Munich, Germany}
 \address[mpp]{Max Planck Institute for Physics, F\"{o}hringer Ring 6, 80805 M\"unchen, Germany}
 \address[mpip]{Max Planck Institute for Plasma Physics, Wendelsteinstr. 1, 17491 Greifswald, Germany}
 \address[nsu]{Novosibirsk State University, 630090, Novosibirsk, Russia}
 \address[ral]{STFC Rutherford Appleton Laboratory, Didcot, OX11 0QX, UK}
\address[epfl]{Swiss Plasma Center, EPFL, 1015 Lausanne, Switzerland}
 \address[tum]{Technische Universit\"at M\"unchen (TUM), Arcisstrasse 21, D-80333 Munich, Germany}
 \address[triumf]{TRIUMF, 4004 Wesbrook Mall, Vancouver V6T2A3, Canada}
 \address[ucl]{UCL, Gower Street, London WC1E 6BT, UK}
 \address[liv]{University of Liverpool, Liverpool L69 7ZE, UK}
 \address[manc]{University of Manchester, Manchester M13 9PL, UK}
 \address[oslo]{University of Oslo, 0316 Oslo, Norway}
 \address[ox]{University of Oxford, Oxford, OX1 2JD, UK}
 \address[strath]{University of Strathclyde, 16 Richmond Street, Glasgow G1 1XQ, UK}
 \address[uv]{University of Victoria, 3800 Finnerty Rd, Victoria, Canada}

\begin{abstract}
The Advanced Proton Driven Plasma Wakefield Acceleration Experiment (AWAKE) aims at studying plasma wakefield generation and electron acceleration driven by proton bunches. It is a proof-of-principle R$\&$D experiment at CERN and the world's first proton driven plasma wakefield acceleration experiment. The AWAKE experiment will be installed in the former CNGS facility and uses the 400~GeV/c proton beam bunches from the SPS. The first experiments will focus on the self-modulation instability of the long (rms $\sim$~12~cm) proton bunch in the plasma. These experiments are planned for the end of 2016. Later, in 2017/2018, low energy ($\sim$~15~MeV) electrons will be externally injected to sample the wakefields and be accelerated beyond 1~GeV. The main goals of the experiment will be summarized. A summary of the AWAKE design and construction status will be presented.
\end{abstract}


\end{frontmatter}

\section{Introduction}
\begin{figure}[htb]
\centering
\includegraphics[width=85.4mm]{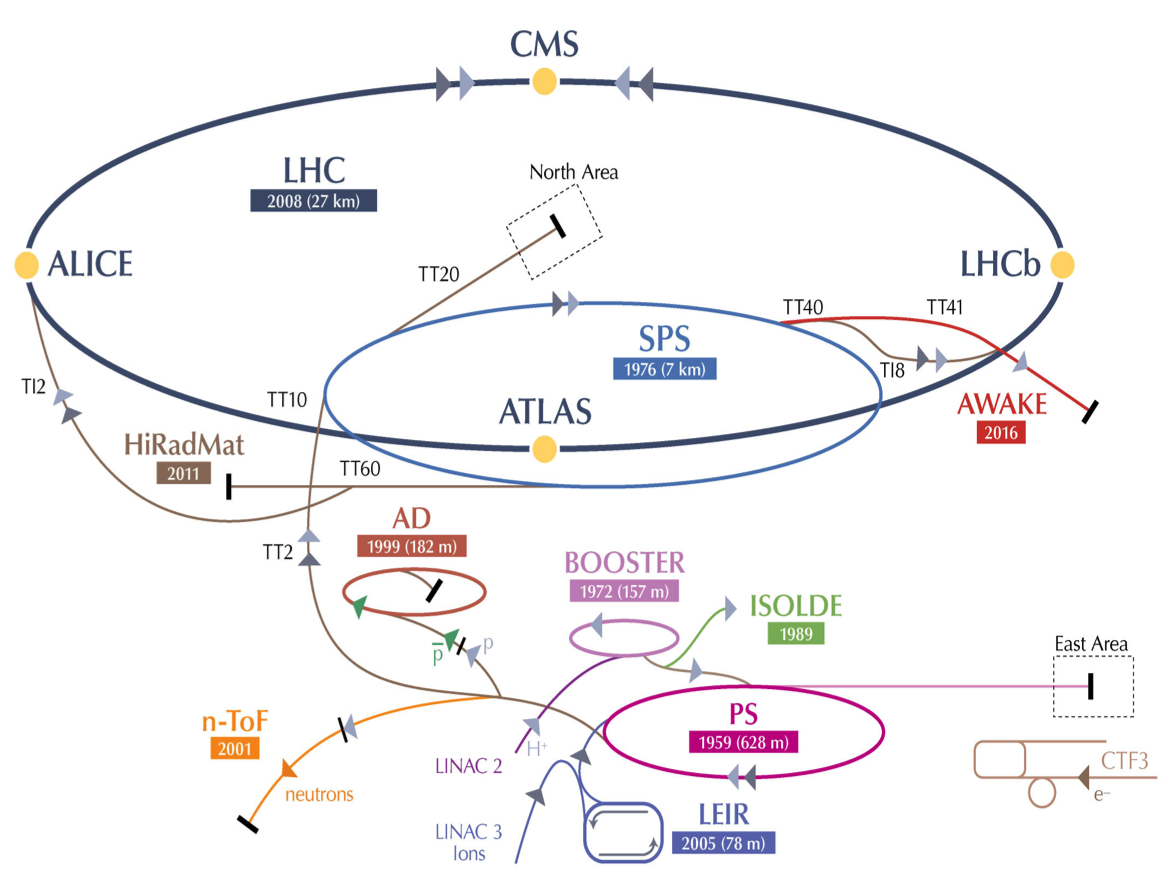}
\caption{CERN accelerator complex.}\label{f-cern-acc}
\end{figure}

AWAKE is a proof-of-concept acceleration experiment with the aim to inform a design for high energy frontier particle accelerators and is currently being built at CERN~\cite{AwakeDR, awake-paper1}.
The AWAKE experiment is the world's first proton driven plasma wakefield acceleration experiment, which will use a high-energy proton bunch to drive a plasma wakefield for electron beam acceleration.
A 400~GeV/c proton beam will be extracted from the CERN Super Proton Synchrotron, SPS, and utilized as a drive beam for wakefields in a 10~m long plasma cell to accelerate electrons with amplitudes up to the GV/m level.
Fig. \ref{f-cern-acc} shows the AWAKE facility in the CERN accelerator complex.
In order to drive the plasma wakefields efficiently, the length of the drive bunch has to be on the order of the plasma wavelength $\lambda_{pe}$, which corresponds to $\approx$1~mm for the plasma density used in AWAKE ($10^{14} - 10^{15}$ electrons/cm$^{3}$).
The proton beam for AWAKE has a bunch length of $\sigma_z = 12~$cm, therefore the experiment relies on the self-modulation instability (SMI) \cite{smi}, which modulates the proton driver at the plasma wavelength in the first few meters of plasma.
The SMI is a transverse instability that arises from the interplay between transverse components of the plasma wakefields and the wakefields being driven by regions of different bunch densities.
The modulation period s$\cong \lambda_{pe}$ and the modulated bunch resonantly drives the plasma wakefields.
The occurence of the SMI can be detected by characterizing the longitudinal structure of the proton beam when exiting the plasma cell.

In the AWAKE master schedule, the experiment to obtain evidence for the SMI  corresponds to Phase 1, and is expected to start by the end of 2016.
In Phase 2, AWAKE aims at the first demonstration of proton-driven plasma wakefield acceleration of an electron witness beam; this programme is planned to start by the end of 2017.
At a later phase it is foreseen to have two plasma cells in order to separate the modulation of the proton bunch from the acceleration stage. Simulations \cite{allen-kostya} show that this would optimize the acceleration of external electrons and reach even higher gradients.

%
\subsection{Baseline Design}
In the baseline design of AWAKE at CERN, an LHC-type proton bunch of 400~GeV/c (with an intensity of $\approx 3\times 10^{11}$~protons/bunch) will be extracted from the CERN SPS and sent along the 750~m long proton beam line towards a plasma cell.
The AWAKE facility is installed in the area, which was previously used for the CERN Neutrinos to Gran Sasso facility (CNGS)~\cite{gschwendtner2010}.
The proton beam will be focused to $\sigma_{x,y} = 200~\mu$m near the entrance of the 10~m long rubidium vapor plasma cell with an adjustable density in the $10^{14}$ to $10^{15}$~electrons/cm$^3$ range.
When the proton bunch, with an r.m.s. bunch length of $\sigma_{z} ~ = 12$~cm (0.4~ns), enters the plasma cell, it undergoes the SMI. The effective length and period of the modulated beam is set by the plasma wavelength (for AWAKE, typically $\lambda_{pe} = 1$~mm). A high power ($\approx$~4.5~TW) laser pulse, co-propagating and co-axial with the proton beam, will be used to ionize the neutral gas in the plasma cell and also to generate the seed of the proton bunch self-modulation.
An electron beam of $1.2\times 10^{9}$~electrons, which will be injected with $10 - 20$~MeV/c into the plasma cell, serves as a witness beam and will be accelerated in the wake of the modulated proton bunch.
Several diagnostic tools will be installed downstream of the plasma cell to measure the proton bunch self-modulation effects and the accelerated electron bunch properties.
\begin{figure}[htb]
\centering
\includegraphics[width=85.4mm]{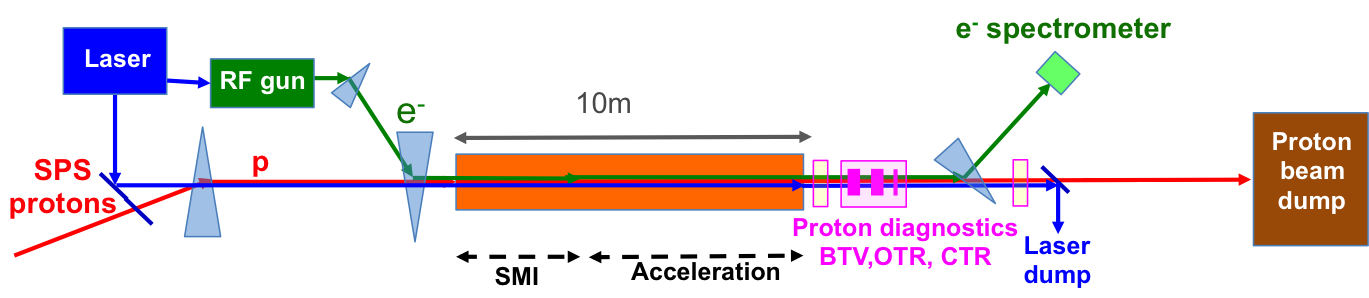}
\caption{Baseline design of the AWAKE experiment.}\label{f-awake-sketch}
\end{figure}
Fig.~\ref{f-awake-sketch} shows the baseline design of the AWAKE experiment. The baseline parameters of the proton beam, laser and plasma cell are summarized in Table~\ref{parameters1}; those of the electron beam in Table~\ref{parameters2}.

\begin{table*}
\centering
\caption{AWAKE proton, laser beam and plasma parameters}
\label{parameters1}
\begin{tabular}{lc}
\hline \hline
Parameter & Baseline  \\
\hline \hline
Proton beam & \\
\hline
Beam momentum & 400 GeV/c  \\
Protons/bunch & 3$\times 10^{11}$   \\
Bunch extraction frequency & 0.03 Hz (ultimate: 0.14~Hz) \\
Bunch length ($\sigma$) & 0.4~ns \\
Bunch size at plasma entrance ($\sigma_{x,y}$) & 200~$\mu$m \\
Normal. emittance (RMS) & 3.5 mm~mrad \\
Relative energy spread ($\Delta$p/p) & 0.035\% \\
Beta function ($\beta^{*}_{x,y}$) & 4.9~m \\
Dispersion ($D^{*}_{x,y}$) & 0 \\
\hline
Laser beam to plasma& \\
\hline
Laser type & Fibre titanium: sapphire \\
Pulse wavelength ($L_0$) & 780~nm \\
Pulse length & 100 - 120~fs \\
Laser power & 4.5~TW \\
Focused laser size ($\sigma_{x,y}$) & 1~mm \\
Energy stability (RMS) & $\pm$1.5~\% \\
Repetition rate & 10 Hz \\
\hline
Plasma source & \\
\hline
Plasma type & Laser ionized rubidium vapor \\
Plasma density & $7 \times 10^{14}$cm$^{-3}$ \\
Length & 10~m\\
Plasma radius & $\ge$1~mm \\
Skin depth & 0.2~mm \\
Wavebreaking field $E_0 = mc\omega_{cp}/e$ & 2.54~GV/m \\
\hline \hline
\end{tabular}
\end{table*}
\begin{table*}[ht]
\centering
\caption{AWAKE electron beam parameters}
\label{parameters2}
\begin{tabular}{lcc}
\hline
Parameter & Baseline & Possible Range \\
\hline
Beam energy & 16 MeV & 10-20 MeV \\
Energy spread & 0.5 \% & 0.5 \% \\
Bunch length ($\sigma$) & 4 ps & 0.3-10 ps \\
Beam size at focus ($\sigma$) & 250 $\mu$m & 0.25-1 mm\\
Normalized emittance (RMS) & 2 mm mrad & 0.5-5 mm mrad \\
Charge per bunch & 0.2 nC & 0.1-1 nC \\
\hline
\end{tabular}
\label{beam}
\end{table*}
%
%

\section{The AWAKE Beams: Proton, Electron, Laser}

\subsection{Proton Beam Line}
\label{}
For the main part of the beam line, the 750~m long CNGS transfer line can be reused without major changes. However, in the last 80~m a chicane has been integrated in order to create space for a mirror of the laser beam line, necessary to merge the ionising laser pulse with the proton beam about 22~m upstream the plasma cell. The proton beam is shifted horizontally by 20~mm at the position of the laser mirror in the present layout~\cite{bracco-ipac14}. Two beam position monitors (BPMs) and a beam loss monitor (BLM) are placed around the merging mirror in order to interlock the SPS extraction in case the mirror is hit by the proton beam.
The synchronization of the two co-propagating beams has to be stable to 100~ps and the transverse pointing accuracy of the proton beam at its focal point is required to be $\leq 100~\mu$m and $\leq 15~\mu$rad, so that the proton trajectory is coaxial with the laser over the full length of the plasma cell. Beam position monitors and screens (BTVs) in combination with a streak camera close to the plasma cell allow the overlap and synchronization of the beams to be measured. Optics simulations predict a $1\sigma$ spot size of $210~ \mu$m at the focal point in agreement with the experiment requirements (see Table~\ref{parameters1}).

\subsection{Electron Source}
\label{}
\begin{figure}[htb]
\centering
\includegraphics[width=80.4mm]{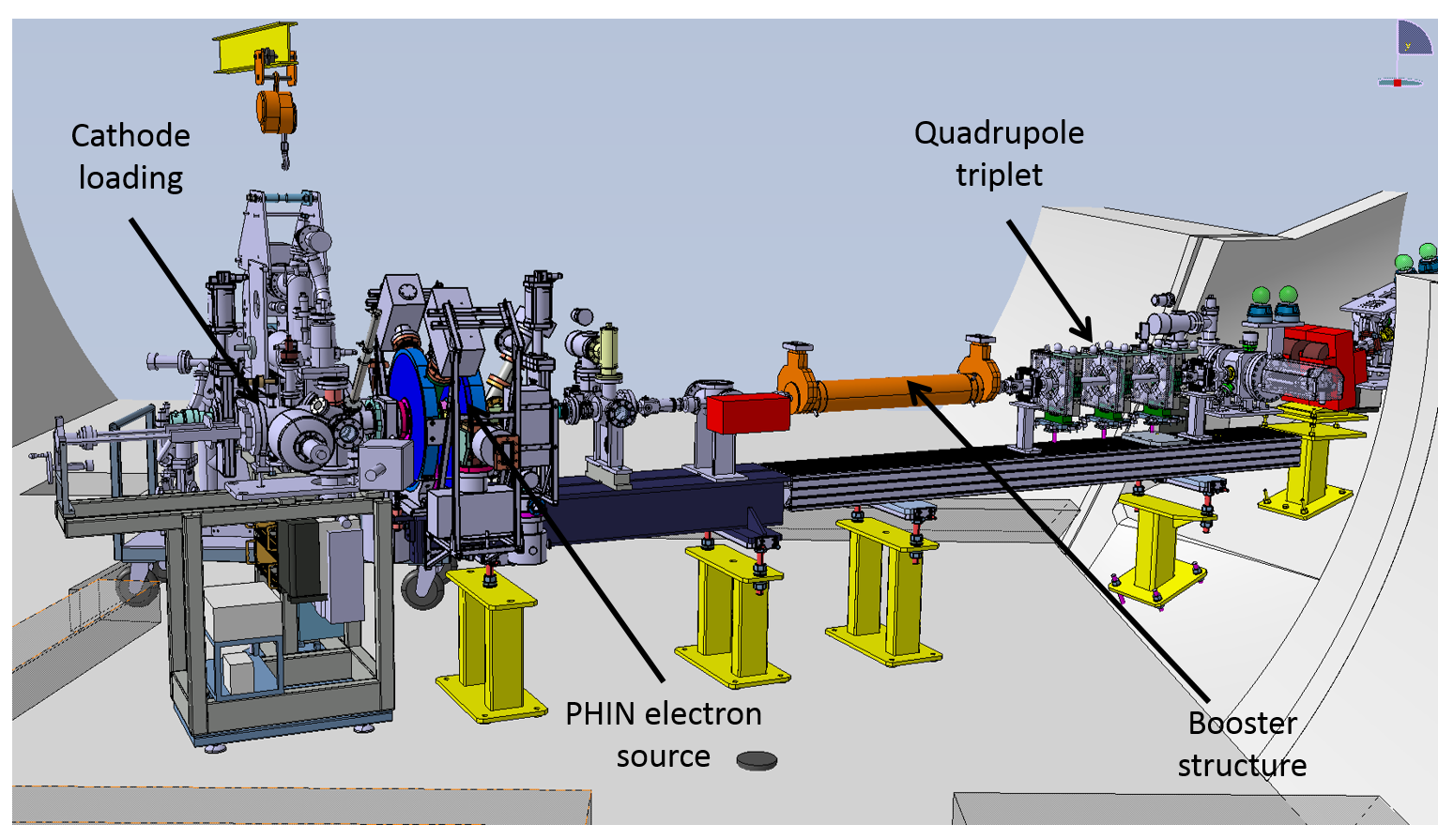}
\caption{Electron source and accelerating structure layout.}\label{f-phin}
\end{figure}

The electron source for AWAKE consists of a 2.5 cell RF-gun and a one meter long booster structure both at 3~GHz (see Fig.~\ref{f-phin}). The electron beam is produced via photo-emission by illuminating a cathode with a frequency quadrupled laser pulse which is derived from the main drive laser for the plasma. The wavelength used in the photo injector will be 262~nm. The baseline will use copper cathodes with a quantum efficiency of $Q_e \approx 10^{-4}$.
However, thanks to the integration of a load lock system, which allows transferring cathodes under ultra high vacuum, different cathodes could be used: e.g. $Cs_2Te$ with a quantum efficiency of $Q_e \approx 10^{-2}$.
A 30 cell travelling wave structure was designed to boost the energy with a constant gradient of 15~MV/m up to a total electron energy of 20~MeV.
The RF-gun and the booster are powered by a single klystron delivering about 30~MW. The operation mode will be single bunch with a maximum repetition rate of 10~Hz.
In addition the electron source is equipped with  BPMs with a resolution of 50~$\mu$m to control the beam position, a fast current transformer with a resolution of 10~pC, a Faraday Cup, and two emittance measurement stations. Details of the electron source are described in~\cite{kevin-steffen-eaac2015}.
The PHIN Photo-Injector built for CTF3~\cite{CTF3}  will be used as the RF-gun for AWAKE. The RF power source will also be recuperated from the CTF3.
\subsection{Electron Beam Line} \label{electron-line}
\begin{figure}[htb]
\centering
\includegraphics[width=55.4mm]{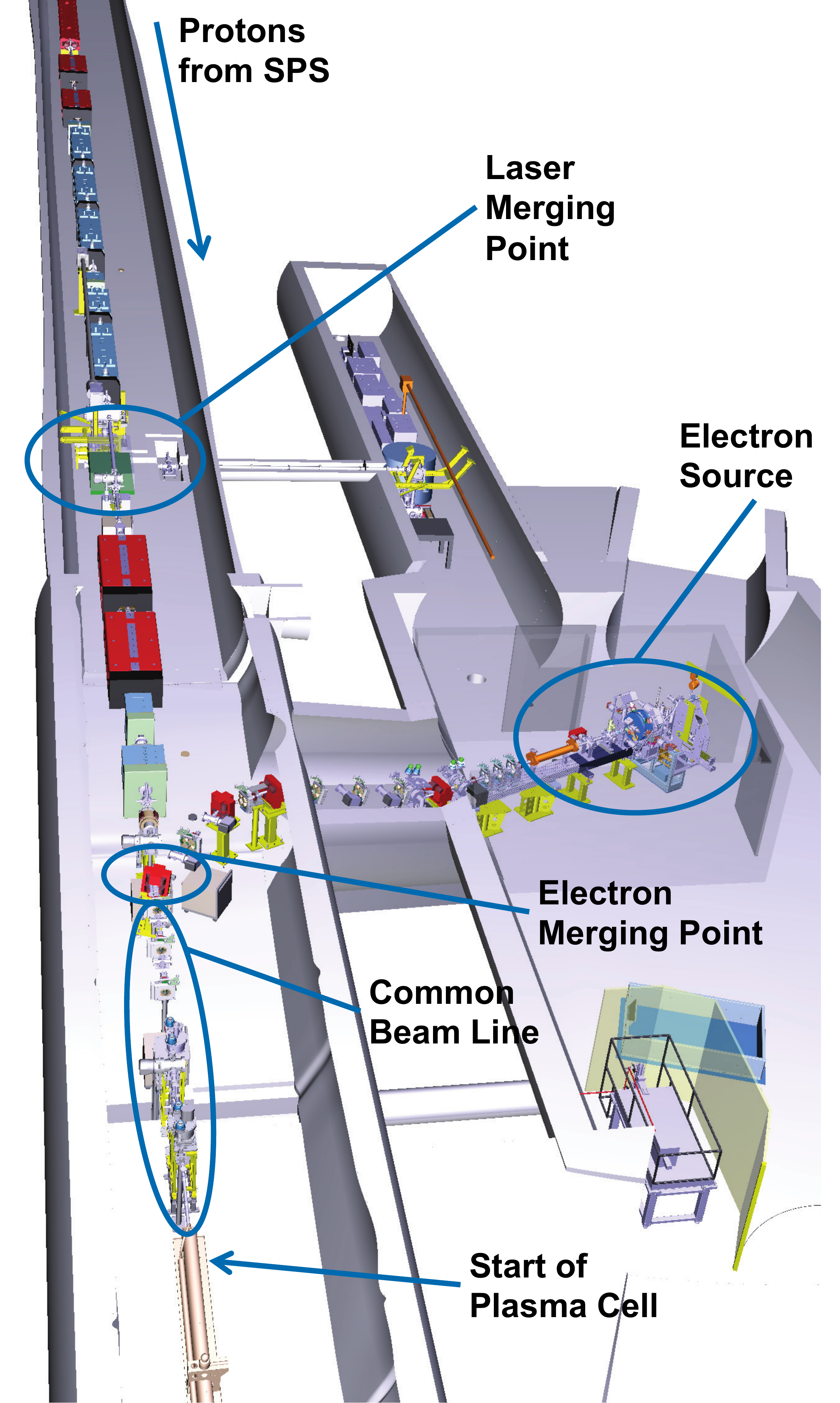}
\caption{The integration of the AWAKE beam lines (proton, electron and laser) in the AWAKE facility (formerly CNGS).}\label{f-sketch-beamlines}
\end{figure}

\begin{figure}[htb]
\centering
\includegraphics[width=88.4mm]{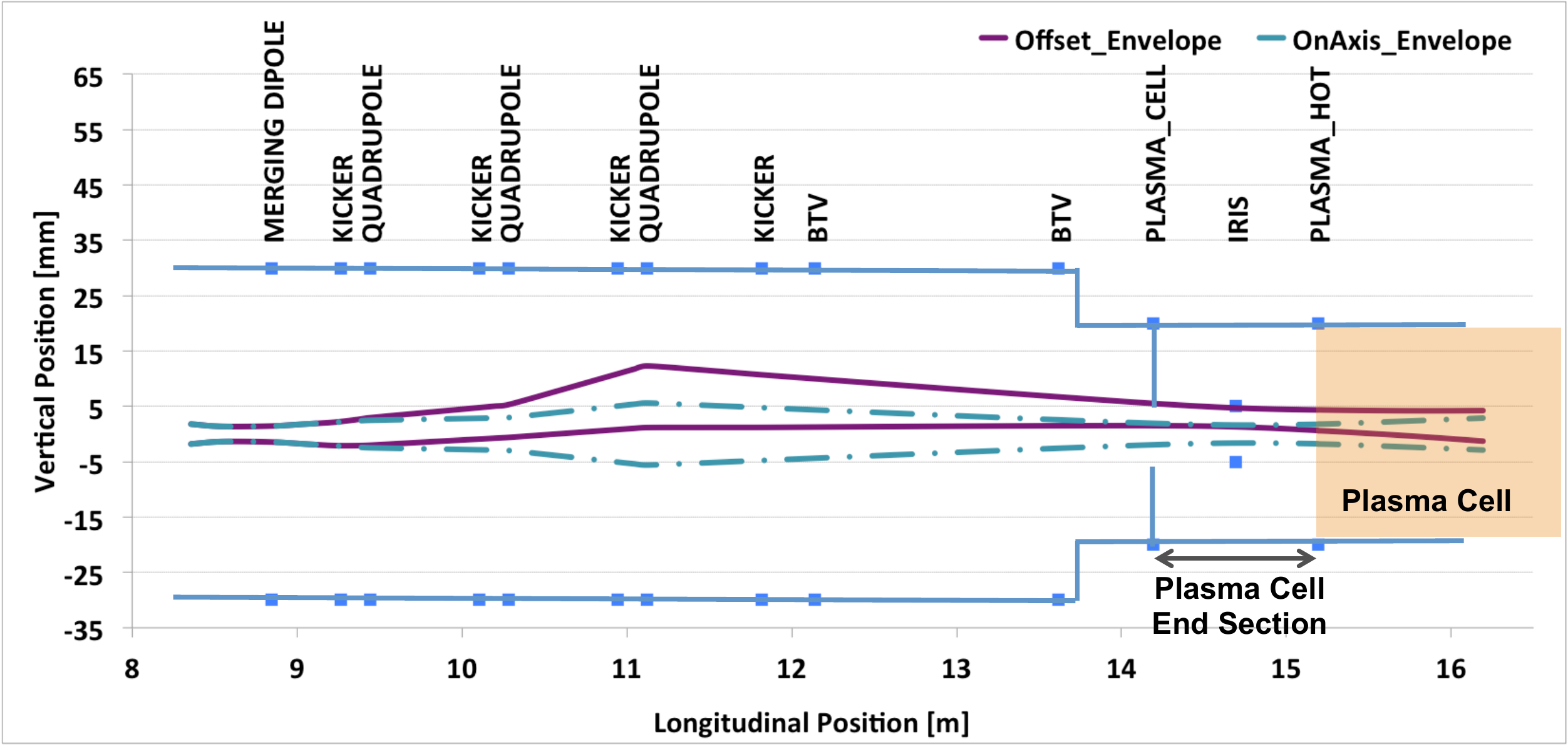}
\caption{The 3$\sigma$ envelope of the electron beam in the common beam line, in case the electron beam is coaxial with the proton beam (``OnAxis") or with an offset to the proton beam axis (``Offset"). The position and free aperture of the main beam line elements are represented by blue squares.}\label{f-e-injection}
\end{figure}

The electron source is installed in an adjacent room 1.16~m below the level of the proton beam line.
The electron transfer line consists of an achromatic dog-leg to raise the electron beam up to the level of the proton line and a part which bends the electron beam horizontally onto the proton beam axis. Details of the beam line design are described in~\cite{janet-eaac2015}.
Fig.~\ref{f-sketch-beamlines} shows the layout of electron beam line, together with the proton and laser beam lines in the AWAKE facility.
Directly after the accelerating structure of the electron source a quadrupole triplet matches the electron beam into the transfer line (see Fig.~\ref{f-phin}).
Another quadrupole triplet just before the plasma is used to focus the beam into the plasma.
Five additional quadrupoles are used to control the dispersion and the beta function.
While the dispersion in the horizontal plane is almost zero along the part of the beam line downstream of the merging dipole (common line with the protons), the dispersion in the vertical plane is not closed due to the vertical kick given by the tilted dipole, which merges the electron beam onto the proton beam axis.
However, the final focusing system matches the beta functions and dispersion to the required $1\sigma$ spot size of $\le 250\mu$m at the focal point in both planes.
Longitudinally the focal point is set at an iris (orifice) with a free aperture of 10~mm about 0.5~m upstream of the plasma cell.
The present optics provides the possibility to shift the focal point up to 0.8~m into the plasma cell without significant changes of the beam spot size.
Ten kickers  (correctors) along the electron beam line compensate systematic alignment and field errors and a shot-to-shot stability of $\pm 100~\mu$m is predicted for a current fluctuation of 0.01$\%$ in the power converters.
In the common beam line upstream the plasma cell the proton, electron and laser beam are travelling coaxially.
Studies on the proton induced wakefields on the beam pipe walls and their effect on the electrons show that the influence of the proton beam wakefields on the electrons is negligible \cite{ulrich-ipac15}.
However, direct beam-beam effects show that the electron beam emittance blows up \cite{ulrich-indico}.
In addition to the beam-beam effect, the most efficient electron injection into the plasma wake fields must be optimized.
The trapping of the electrons in the plasma was analysed with respect to the transverse vertical position $y$ and angle $y'$ at injection \cite{alexey-indico}.
Combining the results of these studies show  a possible optimized injection scheme; the electron beam is offset in the common beam line and this offset is kept also at the injection point (see Fig. \ref{f-e-injection}).
Adding a kicker magnet close to the plasma cell and focusing the beam into the iris allows to inject the electrons into the plasma wakefield with an offset of up to 3.25~mm and an angle between 0 and 8~mrad \cite{janet-eaac2015}.

\subsection{Laser Beam Line}
The laser system is housed in a dust-free, temperature-stabilized area (class 4) and includes the laser, pulse compressor and laser beam transport optics~\cite{status-spsc2015}.
The laser beam line to the plasma cell starts at the output of the optical compressor in the laser lab and is transported via a newly drilled laser core. The laser beam line is enclosed in a vacuum system, which is attached to the compressor's vacuum chamber and to the proton beam line vacuum system at the merging point, which is at a vacuum level of $10^{-7}$~mbar.
The deflection of the laser beam will be performed with dielectric mirrors held by motorized mirror mounts, installed in the vacuum system. The size of the laser beam and the focusing spatial phase will be controlled by a dedicated telescope just before the pulse compressor. The focal distance is to be adjustable in the range between 35 and 45~m.

A diagnostic beam line will be installed in the proton tunnel for measuring the beam properties of a low energy replica of the ionizing beam in exactly the distance which corresponds to the plasma cell location.

The laser beam for the electron gun will be taken from the second output of the base version of the laser system, further amplified to $\approx$30~mJ, compressed to 300~fs using an in-air compressor, frequency converted to 262~nm via a Third Harmonics Generation and stretched to the desired pulse length of 10~ps.

\subsection{Low-Level RF and Synchronization}

The proton, electron and the high power laser pulse have to arrive simultaneously in the rubidium plasma cell.
The proton bunch is extracted from the SPS about every 30 seconds and must be synchronized with the AWAKE laser and the electron beam pulsing at a repetition rate of 10 Hz.
The latter is directly generated using a photocathode triggered by part of the laser light, but the exact time of arrival in the plasma cell still depends on the phase of the RF in the accelerating structure.
Each beam requires RF signals at characteristic frequencies: 6 GHz, 88.2 MHz and 10 Hz for the synchronization of the laser pulse, 400.8 MHz and 8.7 kHz for the SPS, as well as 3 GHz to drive the accelerating structure of the electron beam~\cite{marzia-ipac15}.
A low-level RF system and distribution has been designed to generate all signals derived from a common reference.
Additionally precision triggers, synchronous with the arrival of the beams, will be distributed to beam instrumentation equipment to measure the synchronization.
Phase drifts of the optical fibers transporting the RF signals for the synchronization of the SPS with AWAKE will be actively compensated by newly developed hardware, which is essential to achieve a link stability of the order of 1~ps.
\section{Plasma Source} \label{plasmacell}
AWAKE will use a rubidium vapor source~\cite{erdem-eaac2015} ionized by a short laser pulse (see Table \ref{parameters1}). Rubidium plasma consists of heavy ions, that  mitigate plasma ions motion effects.
The plasma cell is 10~m long and has a diameter of 4~cm.
The density uniformity is achieved by imposing a uniform temperature (within 0.2~\%) along the source.
For that purpose a heat exchanger with sufficient heat carrying fluid flow is used. Synthetic oil is circulated inside a thermal insulation around the tube containing the rubidium vapor.
The oil temperature can be stabilized to $\pm$0.05$ ^{\circ}$ C.
A threshold ionization process for the first Rb electron is used to turn the uniform neutral density into a uniform plasma density.
The ionization potential is very low, $\Phi_{Rb}$=4.177~eV, as is the intensity threshold for over the barrier ionization (OBI), $I_{Ioniz}\approx 1.7 \times 10^{12}$~W/cm$^2$.

At the two ends of the 10~m long plasma cell fast valves were foreseen originally, with the fast valves open only to let the proton, laser and electron beam pass.
However, gasdynamic simulations of the Rb vapor flow showed that the fast valves were not fast enough (1-3~m) to ensure a short enough density ramp:
for efficient electron trapping one needs the density ramp shorter than $\approx$10~cm.
To meet this requirement the vapor cell ends will now have a continuous flow through orifices at each end~\cite{status-spsc2015}.
The Rb sources should be placed as close as possible to the orifices to minimize the density ramp length.
Thus there is continuous flow of Rb from the sources to the plasma cell and afterwards from the plasma cell to the expansion volumes through the orifices (10~mm diameter).
The walls of the expansion volumes should be cold enough ($39 ^{\circ}$~C, the melting temperature of Rb) to condense all Rb atoms.
The density gradient in the plasma cell will be controlled by the temperature difference of the Rb sources at each end of the plasma cell.
In order to control a density gradient of 0.5\% to 1\% with at least 50\% precision the relative reservoir temperatures must be controlled with a precision of 0.1$^{\circ}$~C or better.
\section{Electron Injection}
In order to optimize the electron acceleration in the plasma various schemes for electron injection have been investigated.
The history of the evolution on the optimized electron injection for AWAKE is described in \cite{path}.
In the first experimental phase the electron bunch will be at least one plasma period long in order to avoid exact phasing with the proton bunch modulation and thus cover several modulation cycles.
Once the SMI is better understood and optimal parameters are found, it is planned to inject short electron bunches at the desired phase.

\subsection{Oblique Injection}
At the plasma entrance the plasma density increases smoothly from zero to the baseline density of $7 \times 10^{14}$electrons/cm$^{3}$.
Electrons initially propagating along the proton beam axis are not trapped by the plasma wave in case the plasma density increases over a too long distance.
The effect is similar to the plasma lens effect \cite{plasma-lens} and is explained in detail in ~\cite{path}.
For the parameters of the AWAKE experiment, a transition region of 10~cm length is sufficient to defocus the electrons.

To shorten the transition area, the plasma cell ends are designed with a continuous flow through orifices as described in section \ref{plasmacell}.
With this design the defocusing region is on the order of 15~cm.
This distance is still sufficient to deliver a radial momentum of about 0.5~MeV/c to the electrons thus preventing their trapping by the plasma wave.
But fortunately the defocusing region does not extend beyond the radial plasma boundary~\cite{path}.
So the electrons that are outside the ionized area in the plasma transition region propagate freely and some of them can even receive a small focusing push of several mrad.

\begin{figure}[htb]
\centering
\includegraphics[width=85.4mm]{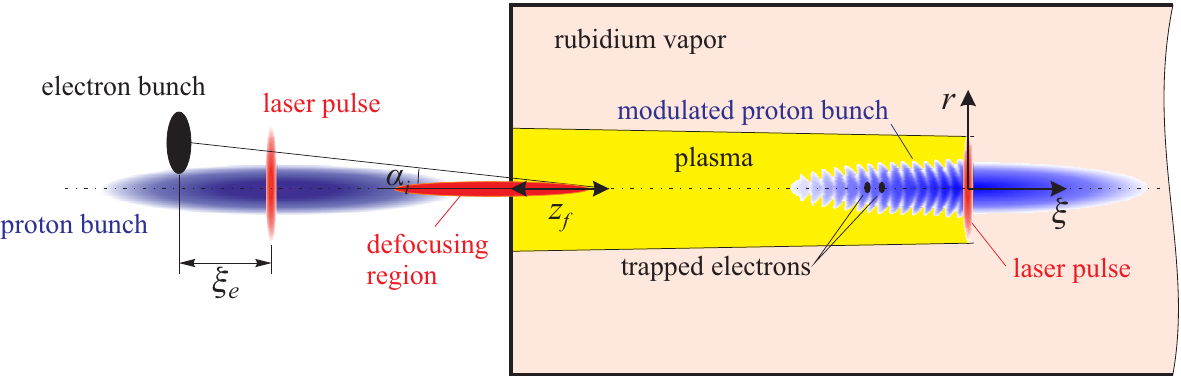}
\caption{The oblique injection of the electrons into the plasma.}\label{f-oblique}
\end{figure}

Applying an oblique electron injection as shown in Fig.~\ref{f-oblique} mitigates the loss of the electrons at the plasma density transition region:
the electrons have a small radial offset with respect to the proton beam upstream the plasma cell and are injected with a small angle $\alpha_i$ into the plasma.
In this way they approach the axis in the region of already constant plasma density and therefore can get trapped into the established plasma wave.
The optimum values for the oblique injection scheme found in simulations are~\cite{path}: electron delay $\xi_e=11.5$\,cm, injection angle $\alpha_i = 2.8$\,mrad, and focusing point $z_f = 140$\,cm.
These values are compatible with the range of settings for the electron beam line (see section \ref{electron-line}) and therefore no changes in the facility design are required.
%
%
\subsection{Density Gradient along the Plasma}
\begin{figure}[htb]\centering
\includegraphics[width=70.4mm]{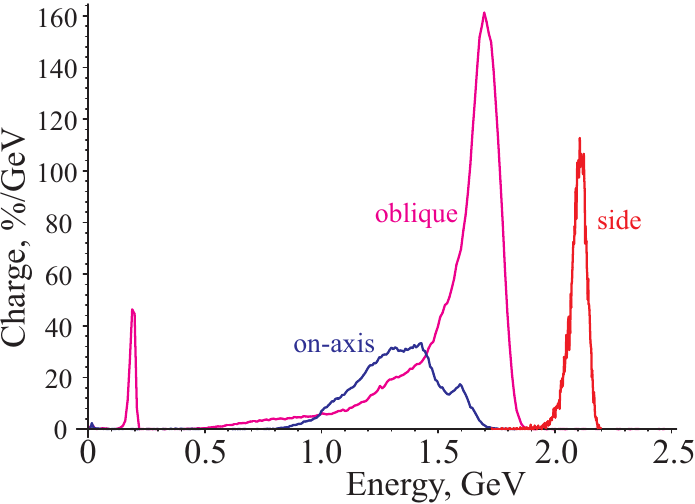}
\caption{Final energy spectra of electrons in cases of side, on-axis, and oblique injection methods.   Beam loading is taken into account.}\label{f-espectra}
\end{figure}

With the plasma cell design as described in section \ref{plasmacell} a plasma density gradient of several percent along the 10~m long plasma cell can be created.
If the gradient is positive the resulting change of the wakefield structure optimizes the electron acceleration.
Details of these studies are described in~\cite{alexey-eaac2015}.
Fig.~\ref{f-espectra} shows the final energy of the electrons after passing through the 10~m long plasma cell for oblique, on-axis and side-injection.
(In side-injection, the electrons would be injected into the plasma cell only after the SMI has developed.
This has turned out to be technically very challenging and was abandoned for the first phases of the experiment~\cite{bracco-ipac14}.)
The results show that with oblique injection, realistic plasma boundaries at both ends and the linear growth of the plasma density by 1~\% over 10~m, about 40~\% of the injected electrons are trapped and accelerated to $\approx$1.8~GeV after 10~m.
\section{Diagnostics}

\subsection{Direct SMI Measurements}
The direct SMI diagnostic tools are based on transition radiation measurements~\cite{patric-ipac2015}.

The optical transition radiation (OTR) is prompt and its time structure reflects that of the bunch charge density hitting the radiator foil placed in the beam path.
For AWAKE the plasma and modulation frequency range is 100-300~GHz.
The time structure of the light and the proton bunch can be characterized using a $\cong$1~ps resolution streak camera.
The ability of the streak camera to detect the ps modulation of the light signal was tested using beating laser beams; the results show that the period of the modulation can be measured for frequencies of up to 300~GHz~\cite{status-spsc2015}.

Coherent transition radiation (CTR) diagnostics uses time-resolved and heterodyne frequency measurements to determine the frequency and possibly amplitude of the SMI.
The CTR has a radially polarized electric field and has a maximum emission at a certain angle that depends on the particular experimental conditions of the plasma density. This makes it difficult to couple into the fundamental mode of circular or rectangular wave-guides, therefore using quasi-optical propagation to handle the CTR beam at least in the vicinity of the interaction point is considered.
The CTR can be detected when focused onto a pyro-detector or to couple part of a beam into a horn antenna and then detect it with the help of Schottky diodes.

\subsection{Indirect SMI Measurement}
Protons in the defocusing phase of the wakefields exit the plasma close to the plasma center and appear as a narrow core on transverse profiles downstream the plasma.
In \cite{marlene-eaac2015} it is shown that the defocusing angle is on the order of 1~mrad for the 400~GeV/c protons.
For the indirect SMI measurements two beam-imaging screens (BTVs) at a distance of $\approx8$~m will be inserted downstream the plasma cell in order to measure the transverse bunch shape and the beam size.
Detailed studies of the screen material showed that a 1~mm thick Chromox-6 (Al$_2$O$_3$:CrO$_2$) scintillator with a hole or an OTR material at the beam centre can measure the beam shape and produce enough light, with only minimal interference of the proton beam.
A defocused beam edge resolution of 0.6~mm can be achieved \cite{marlene-eaac2015}.

\subsection{Electron Spectrometer}
The electron spectrometer system~\cite{lawrence-ipac2015} consists of a C-shaped dipole providing a 1.5~T field to separate the electrons from the proton beam and disperse them in energy onto a scintillating screen (baseline is gadolinium oxysulfide).
The spectrometer system also includes a quadrupole doublet in a point-to-point imaging configuration, to focus the beam exiting the plasma onto the spectrometer screen and increase the energy resolution.
An optical line will transport the screen light to a CCD camera.
Radiation protection calculations showed that the CCD camera needs to be located at a position 17~m away from the screen in order to avoid radiation damage.
Calculations show that by transporting the light to the camera using a series of mirrors and a suitable commercially available lens a peak signal to noise ratio of at least 120 is achieved at the expected electron acceleration capture efficiency.
\section{The AWAKE Underground Facility and Safety Aspects}
Modifying the CNGS underground high radiation area for AWAKE beam requires challenging modifications in a complex experimental area.
AWAKE services and infrastructure must be integrated within the existing radiation facility and at the same time be designed and installed to keep the radiation dose to personnel as low as possible.
Shielding was installed and CNGS elements dismantled or exchanged in order to turn the high-radiation, no-access CNGS facility into a supervised radiation area with safe regular access.
Fig.~\ref{f-sketch-beamlines} shows the integration of the AWAKE beam lines in the CNGS facility; the diagnostics is downstream of the plasma cell and not seen in the figure.

Fire safety equipment and the access procedures must be modified following extensive fire risk assessments, in order to deal with the specific AWAKE access needs (a stop-and-go proof-of-principle experiment) and properties of the new equipment.
Because of the high radiation, materials in CNGS were limited to concrete, iron, graphite etc. leading to a very low fuel load of the area. The much lower radiation level in AWAKE means that racks, electronics and other equipment with a non-negligible fuel load are still installed in an underground area that is 1~km distance from the nearest exit.
In order to guarantee fire safety, a new and more complex fire zone layout was created.
Fire resistant walls and doors will ensure that at least two evacuation paths are available for every AWAKE area.
Fire risk assessment were performed for the regions with the higher fuel loads (e.g. the oil-based Klystron for the electron source), fire intervention and evacuation exercises are organised and the fire safety and evacuation information will be part of a dedicated AWAKE safety course, obligatory for every person wanting to enter the area.

Radiation protection calculations have been performed using FLUKA Monte Carlo code~\cite{fluka1, fluka2} to study the radiation environment in the AWAKE facility.
The results show that access to the AWAKE experimental area has to be prohibited during the proton beam operation as the prompt dose equivalent rate exceeds 100 mSv/h. As a consequence, the experiment and equipment must be remotely controlled.
To allow partial access to the laser room and the klystron area during the operation of the electron beam, an appropriate shielding wall around the electron gun has to be designed.
During electron operation radiation levels in accessible areas will be continuously monitored.
Moreover, a dedicated air management system of the area including an airborne radioactivity monitor has to be installed to guarantee safe access conditions after proton beam operation.
\section{Summary}
AWAKE is a proof-of-principle accelerator R\&D experiment currently being built at CERN. It is the first proton-driven wakefield acceleration experiment worldwide, with the aim to provide a design for a particle physics frontier accelerator at the TeV scale.
The installation of the AWAKE experiment is advancing well. Hardware and beam commissioning is planned for the first half of 2016.
The physics of the self-modulation instability as a function of the plasma and proton beam properties will be studied starting by the end of 2016.
The longitudinal accelerating wakefield will be probed with externally injected electrons starting by the end of 2017.
At a later stage it is foreseen to have two plasma cells in order to separate the modulation of the proton bunch from the acceleration stage and to reach even higher gradients.
\section{Acknowledgment}
This work was supported in parts by: EU FP7 EuCARD-2, Grant Agreement 312453 (WP13, ANAC2); and EPSRC and STFC, United Kingdom.
The contribution of Novosibirsk team to this work is supported by The Russian Science Foundation, grant No.~14-12-00043.
The AWAKE collaboration acknowledges the support of CERN, Max Planck Society, DESY, Hamburg and the Alexander von Humboldt Stiftung.

\bibliographystyle{elsarticle-num-names}
\bibliography{bibl}

\begin{thebibliography}{00}

\bibitem{AwakeDR} A.~Caldwell et al., AWAKE Design Report, A Proton-Driven Plasma Wakefield Acceleration Experiment at CERN, Internal Note CERN-SPSC-2013-013, CERN, Geneva, Switzerland (2013).

\bibitem{awake-paper1} R. Assmann et al. (AWAKE Collaboration), Proton-driven plasma wakefield acceleration: a path to the future of high-energy particle physics, Plasma Phys. Control Fusion 56, 084013 (2014).

\bibitem{smi} N.~Kumar, A.~Pukhov and K.~Lotov, Self-Modulation Instability of a Long Proton Bunch in Plasmas, Phys. Rev. Lett. 104 255003 (2010).

\bibitem{allen-kostya} A.~Caldwell and K.~Lotov, Plasma wakefield acceleration with a modulated proton bunch, Physics of Plasma, 18, 103101 (2011).

\bibitem{gschwendtner2010} E.~Gschwendtner et al., Performance and Operational Experience of the CNGS Facility, Proceedings of the 1st International Particle Accelerator Conference, IPAC2010, Kyoto, Japan, THPEC046, (2010).

\bibitem{bracco-ipac14} C.~Bracco et al., Proceedings of IPAC2014, Dresden, Germany (2014).

\bibitem{kevin-steffen-eaac2015} K.~Pepitone, S.~Doebert et al., The electron accelerator for the AWAKE experiment at CERN, EAAC2015, Elba, Italy, these NIM proceedings, (2015).

\bibitem{CTF3} G.~Geschonke et al., CTF3 Design Report, Tech.Rep. CTF3-Note-2002-047, CERN, Geneva (2002).

\bibitem{janet-eaac2015} J.~Schmidt et al, Status of the Proton and Electron Transfer Lines for the AWAKE Experiment at CERN, EAAC2015, Elba, Italy, these NIM proceedings, (2015).

\bibitem{ulrich-ipac15} U.~Dorda et al., Simulations of Electron-Proton Beam Interaction before Plasma in the AWAKE Experiment, IPAC15, Richmond, USA (2015).

\bibitem{ulrich-indico} U.~Dorda et al., Propagation of electron and proton beams before the plasma, https://indico.cern.ch/event/403300/ (2015).

\bibitem{alexey-indico} A.~Petrenko et al., Latest electron trapping simulation results, https://indico.cern.ch/event/403300/ (2015).

\bibitem{status-spsc2015} A.~Caldwell at al., AWAKE Status Report, CERN-SPSC-2015-032, SPSC-SR-169 (2015).

\bibitem{marzia-ipac15} M.~Bernardini et al., AWAKE, a Proof-of-Principle R\&D Experiment at CERN, IPAC15, Richmond, USA (2015).

\bibitem{erdem-eaac2015} E.~\"Oz, F. Batsch, P.~Muggli, An acccurate Rb density measurement method for a plasma wakefield accelerator experiment using a novel Rb reservoir, EAAC2015, Elba, Italy, these NIM proceedings, (2015).

\bibitem{path} A.~Caldwell and AWAKE Collaboration, Path to AWAKE: Evolution of the concept, EAAC2015, Elba, Italy, these NIM proceedings, (2015).

\bibitem{plasma-lens} P.~Chen,  A Possible Final Focusing Mechanism for Linear Colliders, Part. Accel. 20 (1987) 171 (1987).

\bibitem{alexey-eaac2015} A.~Petrenko, K.~Lotov, A.~Sosedkin, Numerical Studies of Electron Acceleration Behind Self-Modulating Proton Beam in Plasma with a Density Gradient,  EAAC2015, Elba, Italy, these NIM proceedings, (2015).

\bibitem{patric-ipac2015} P.~Muggli, AWAKE, proton-driven plasma wakefield experiment at CERN, IPAC15, Richmond, USA (2015).

\bibitem{marlene-eaac2015} M.~Turner et al., Indirect Self-Modulation Instability Measurement Concept for the AWAKE Proton Beam, EAAC2015, Elba, Italy, these NIM proceedings, (2015).

\bibitem{lawrence-ipac2015} L.C.~Deacon et al., Development of a spectrometer for proton driven plasma wakefield accelerated electrons at AWAKE, IPAC15, Richmond, USA (2015).

\bibitem{fluka1} G.~Battistoni, S.~Muraro, P.R.~Sala, F.~Cerutti, A.~Ferrari, S.~Roesler, A.~Fasso, J.~Ranft, Proceedings of the Hadronic Shower Simulation Workshop 2006, Fermilab 6-8 September 2006, M.~Albrow, R.~Raja eds., AIP Conference Proceeding 896, 31-49 (2007).

\bibitem{fluka2} A.~Ferrari, P.R.~Sala, A.~Fasso, J.~Ranft, FLUKA: a multi-particle transport code, CERN-2005-10, INFN/TC-05/11, SLAC-R-773 (2005).

\end{thebibliography}


\end{document}